\documentclass[12pt]{article}
\usepackage[colorlinks,linkcolor=blue,citecolor=blue,urlcolor=blue,bookmarks,bookmarksnumbered]{hyperref}
\usepackage{amsmath,amssymb,amsfonts}
\usepackage[paper=letterpaper,margin=1.0in]{geometry}
\usepackage{graphicx,xcolor}
\usepackage{cite}

\newcommand{\be}{\begin{equation}}
\newcommand{\ee}{\end{equation}}
\newcommand{\bea}{\begin{eqnarray}}
\newcommand{\eea}{\end{eqnarray}}
\newcommand{\ba}{\begin{array}}
\newcommand{\ea}{\end{array}}
\newcommand{\ben}{\begin{enumerate}}
\newcommand{\een}{\end{enumerate}}
\newcommand{\bi}{\begin{itemize}}
\newcommand{\ei}{\end{itemize}}
\newcommand{\bc}{\begin{center}}
\newcommand{\ec}{\end{center}}
\newcommand{\bfig}{\begin{figure}}
\newcommand{\efig}{\end{figure}}
\newcommand{\bq}{\begin{quotation}}
\newcommand{\eq}{\end{quotation}}
\newcommand{\bt}{\begin{table}}
\newcommand{\et}{\end{table}}
\newcommand{\btab}{\begin{tabular}}
\newcommand{\etab}{\end{tabular}}
\newcommand{\bs}{\begin{slide}}
\newcommand{\es}{\end{slide}}

\newcommand{\IR}{\mathbb{R}}

\let\ba=\overline

\def\IR{\relax\leavevmode{\rm I\kern-.18em R}}
\def\ZZ{\relax\leavevmode
       \ifmmode\mathchoice
       {\hbox{\sf Z\kern-.4em Z}}
       {\hbox{\sf Z\kern-.4em Z}}
       {\lower.9pt\hbox{\scriptsize\sf Z\kern-.36em Z}}
       {\lower1.2pt\hbox{\tiny\sf Z\kern-.36em Z}}
       \else{\sf Z\kern-.4em Z}\fi}
\def\RR{\relax\leavevmode
       \ifmmode\mathchoice
       {\hbox{\sf R\kern-.4em R}}
       {\hbox{\sf R\kern-.4em R}}
       {\lower.9pt\hbox{\scriptsize\sf R\kern-.36em R}}
       {\lower1.2pt\hbox{\tiny\sf R\kern-.36em R}}
       \else{\sf R\kern-.4em R}\fi}

\def\resetby#1#2{\@addtoreset{#2}{#1}}
\def\seceq{\@addtoreset{equation}{section}
              \def\theequation{\thesection.\arabic{equation}}}

\def\Label#1{\label{#1}%
                \smash{\hbox to0pt{\raise1ex\hbox{\tiny[#1]}\hss}}}
\def\noLabels{\let\Label=\label}

\begin{document}

\bc

{\large \bf Snowmass 2022}\\ 
{\large \bf Quantum Gravity and Phenomenology:}\\
{\large \bf Dark Matter, Dark Energy, Vacuum Selection,}\\
{\large \bf Emergent Spacetime, and Wormholes}

\vskip 0.5cm

{\bf Per Berglund$\,{}^{1}$, De-Chang Dai$\,{}^{2}$,  Douglas Edmonds$\,{}^{3}$,  Yang-Hui He$\,{}^{4}$,\\ 
Tristan Hubsch$\,{}^{5}$, Vishnu Jejjala$\,{}^{6}$, Michael J. Kavic$\,{}^{7}$, Djordje Minic$\,{}^{8}$, \\
Samuel Powers$\,{}^{9}$, John H. Simonetti$\,{}^{8}$, Dejan Stojkovic$\,{}^{9}$, Tatsu Takeuchi$\,{}^{8}$}

\ec


\renewcommand{\labelenumi}{${}^{\theenumi}$}

{\it{\small
\begin{enumerate}
\itemsep-0.2em
\item Department of Physics and Astronomy,\\University of New Hampshire, Durham, NH 03824, USA 
\item Center for Gravitation and Cosmology, School of Physical Science and Technology,\\
Yangzhou University, 
Yangzhou City, Jiangsu Province 225002, P. R. China, and\\
CERCA/Department of Physics/Institute for the Science of Origins,\\
Case Western Reserve University, Cleveland, OH 44106-7079, USA
\item Department of Physics, Penn State Hazleton, Hazleton, PA 18202, USA
\item London Institute, Royal Institution of Great Britain, 
Mayfair, London W1S 4BS, UK, and
Department of Mathematics, City, University of London, EC1V 0HB, UK, and\\
Merton College, University of Oxford, OX1 4JD, UK, and\\
School of Physics, NanKai University, Tianjin 300071, P. R. China
\item Department of Physics and Astronomy, Howard University, Washington, DC 20059, USA
\item Mandelstam Institute for Theoretical Physics, School of Physics, NITheCS, and \\
CoE-MaSS, University of the Witwatersrand, Johannesburg, WITS 2050, South Africa
\item Department of Chemistry and Physics, SUNY Old Westbury,
Old Westbury, NY 11568, USA
\item Department of Physics, Virginia Tech, Blacksburg, VA 24061, USA
\item HEPCOS, Department of Physics, SUNY at Buffalo, Buffalo, NY 14260-1500, USA
\end{enumerate}
}}

\renewcommand{\labelenumi}{\theenumi.}


\begin{abstract}
\noindent
We discuss the relevance of quantum gravity to the frontier questions in high energy phenomenology:
the problems of dark matter, dark energy, and vacuum selection 
as well as the problems of emergent spacetime and wormholes.
Dark matter and dark energy phenomenology, and the problem of vacuum selection are discussed within the context of string theory as a model of quantum gravity. 
Emergent spacetime and wormholes are discussed in a more general context of effective theories of quantum gravity.
\end{abstract}



\vspace{0.2cm}

\noindent
{\bf Contact person: Tristan Hubsch, thubsch@howard.edu}

\vspace{0.3cm}

\noindent
{\bf Theory Frontier Topical Groups}:\\
{\bf TF01}: String Theory, quantum gravity, black holes, and\\
{\bf TF09}: Astro-particle physics and cosmology

\newpage
\noindent{\bf Introduction and Summary}: 
The overarching question in high energy physics concerns the union of quantum physics with general relativity, 
\textit{i.e.} how can the two be reconciled into a consistent quantum theory of gravity? 
Phenomenologically, the fundamental quantum description of ordinary matter is captured in
the Standard Model (SM) of particle physics, empirically completed by the discovery of the Higgs boson in 2012 \cite{Particle}. 
Though neutrino masses are sometimes classified as physics beyond the Standard Model (BSM),
no new particles or forces have been discovered that compels us to rethink the SM paradigm.
On the other hand, the observed classical gravitational physics is consistent with Einstein's general theory of relativity \cite{GR}, with the current cosmological observations well described within the $\Lambda$-CDM paradigm \cite{Peebles}.
However, $\Lambda$-CDM necessitates the introduction of dark matter and dark energy, which are concepts
not encompassed in the SM.

Quantum gravity is expected to shed light on the origins of these paradigms in
particle physics (SM) and cosmology ($\Lambda$-CDM), and also lead us to insights on how the two are,
or can be, phenomenologically connected.
Indeed, it has been known for some time that quantum gravity can provide constraints for high energy physics (see \textit{e.g.} \cite{Hawking,Dejan,Vafa,Harlow:2018tng}).
In the summary below, we present research efforts into the relation between quantum gravity and phenomenology in terms of a general formulation of string theory and its applications to dark matter, dark energy, vacuum selection, emergent spacetime, and wormholes.

\noindent
{\bf Constraints on the nature of dark matter}: Various astrophysical observations 
point to intriguing
correlations between dark matter, dark energy, and visible matter on the galactic, and galaxy cluster
scales \cite{Edmonds:2020tug,Edmonds:2020iji}. 
Such correlations \cite{Ho:2010ca, Ho:2011xc, Ho:2012ar, Edmonds:2013hba, Edmonds:2016tio, Ng:2016qvh, Edmonds:2017zhg, Edmonds:2017fce} are yet to be understood from the point of view of the
$\Lambda$-CDM model. 
In this context, a new picture of the origin of dark matter from quantum gravity, in the guise of a general non-commutative formulation of string theory 
\cite{Freidel:2013zga, Freidel:2014qna, Freidel:2015pka, Freidel:2015uug, Freidel:2016pls, Freidel:2017xsi, Freidel:2017wst,Freidel:2017nhg, Freidel:2018apz, Freidel:2019jor, Minic:2020oho, Freidel:2021wpl}, has been
recently proposed \cite{Minic:2020oho,Berglund:2020qcu}. 
Essentially, in this approach, dark matter should be viewed as a dual SM, 
where ``dual'' refers to the
dual spacetime that is found in such a general formulation of string theory. 
This picture leads to a completely new dark matter phenomenology,
 and we would like to understand whether the above mentioned empirical 
scaling relations on galactic and cluster scales \cite{Dai:2017unr}
are consequences of this new view of dark matter.
This will not only sharpen our understanding of the $\Lambda$-CDM model \cite{now1},
but should also provide motivation for further study of various unexplored
correlations in the existing and upcoming simulations of structure formation.

\noindent
{\bf Origin of dark energy in string theory}: 
The current observational discrepancy in the values of the Hubble constant, \textit{i.e.} the $H_0$-tension \cite{Verde:2019ivm}, has placed the $\Lambda$-CDM model under intensified scrutiny.
Recently, we have pointed out that quantum gravity, realized in terms of a general string theory, leads to a dynamical dark energy \cite{Jejjala:2007hh},
which, in turn, may illuminate the $H_0$-tension \cite{Jejjala:2020lhg}. 
This work is a follow-up to the more general discussion of the question ``how to
find de Sitter space in string theory.'' 
Essentially, dark energy is realized as the geometry of the dual spacetime
in the general formulation of string theory \cite{Berglund:2019yjq, Berglund:2019ctg, Berglund:2019pxr}. 
In this context, we have also 
explored certain toy models motivated by string theory,
like the codimension-2 ``deformed stringy cosmic brane" \cite{rBHM1, rBHM2, rBHM3, rBHM4, rBHM5, rBHM6}.

\noindent
{\bf Inflation and its possible relation to dark energy}: Inflation is yet to be
fully understood in the context of a fundamental formulation of quantum gravity, such as string
theory. This question is related to the above important issue of ``how to find de Sitter space in string theory'' 
\cite{Grana:2005jc,Douglas:2006es,Danielsson:2018ztv}.
We have addressed the appearance of de Sitter space in a general non-commutative formulation of
string theory, and we would like to understand 
whether acceptable inflationary 
scenarios can also appear in this formulation  \cite{Berglund:2019yjq, Berglund:2019ctg, Berglund:2019pxr}. 
In particular, we envision the realization of inflation
from a dynamical geometry of the dual spacetime in a general string theory, 
where inflation is  related to
dark energy as in quintessential inflation \cite{Peebles:1998qn}.

\noindent
{\bf The vacuum selection principle in string theory}: One of the fundamental questions of
high energy physics is the origin of the SM, \textit{i.e.} understanding the origins of
its particle content and interactions.
String theory has been concerned
with this problem for more than 30 years \cite{Polchinski:1998rq}.
Recently we have explored a new picture of the vacuum
selection principle \cite{Braun:2005nv, Candelas:2007ac} motivated by  ideas from biophysics, which argue for
the universality of the genetic code based on the idea of horizontal gene transfer \cite{Argyriadis:2019fwb}. We are interested
in applying these ideas in the context of string theory, where the horizontal transfer of information between
different perturbative string vacua is modeled by string interactions. It turns out that the
topological string information about compactifications can be mapped into biological data and the
same mechanism for the universal vacuum selection found in biophysics may be applied in string theory \cite{now3}.
This work also nicely relates to the current efforts to apply the techniques from machine learning and AI in the
study of string vacua \cite{He:2017aed, Carifio:2017bov, He:2020mgx}. 
For example, machine learned Calabi-Yau metrics \cite{Ashmore:2019wzb, Anderson:2020hux, Douglas:2020hpv, Jejjala:2020wcc, Larfors:2021pbb} can be used to find Yukawa couplings in string compactifications to the SM.
Furthermore, we would
also like to understand the role of modular invariance
for modeling of the observed masses and couplings \cite{Feruglio:2017spp}. Specifically, we would like
to explore modular invariance in the realm of the ``stringy cosmic brane" toy
model for dark energy by turning this model into a more realistic scenario that involves SM-like matter,
and ultimately, dark matter and dark energy.
Similarly, we would like to understand constraints of
quantum gravity on various particle masses \cite{Hawking,Dejan}, especially in the context of 
different robust predictions of string theory.
Also, the Higgs sector and the so-called hierarchy problem
can be explored especially in the context of the general
covariant non-commutative formulation of string theory where
various phenomenological consequences of 
a see-saw relation involving the UV and IR cutoffs can be studied
\cite{Berglund:2020qcu, Berglund:2019yjq, Berglund:2019ctg, Berglund:2019pxr}.


\noindent
{\bf Theory and  phenomenology of emergent space-time}: Failure to find a quantum theory of space-time using standard techniques may be an indication that space and time are not fundamental but rather emergent categories. If we relax the demand that the number of dimensions is a fixed number at all length/energy scales, and let the number of effective dimensions increase from UV to IR, we open up a completely new playground to resolve long standing problems in particle physics and cosmology \cite{Anchordoqui:2010er,Anchordoqui:2010hi,Mureika:2011bv,Mureika:2011yy,Stojkovic:2013lga,Stojkovic:2014lha,Afshordi:2014cia,Dai:2014roa,Hao:2014tsa}. In $1+1$ dimensions, the SM hierarchy problem does not exist and  QCD becomes super-renormalizable, while in $2+1$ dimensions the problem of non-renormalizability of general relativity disappears. 
If we open up the fourth spatial dimension at large scales (e.g., the horizon scale),  the infrared divergences   in field theory disappear \cite{Stojkovic:2014lha} and an effective
cosmological constant of the correct magnitude is induced without putting it into
the equations by hand \cite{Anchordoqui:2010er}. 
Some attempts to construct a fundamental theory of an emergent spacetime with such properties already exist \cite{Afshordi:2014cia,Dai:2014roa,Hao:2014tsa},
though more effort is required to obtain such a theory from first principles. 
Also, the prospect of testing such models is very promising, both in the context of colliders \cite{Anchordoqui:2010er,Anchordoqui:2010hi} and gravitational wave observatories \cite{Mureika:2011bv,Mureika:2011yy}. These efforts (which strongly correlate to the research program presented in 
\cite{Freidel:2013zga, Freidel:2014qna, Freidel:2015pka, Freidel:2015uug, Freidel:2016pls, Freidel:2017xsi, Freidel:2017wst,
Freidel:2017nhg, Freidel:2018apz, Freidel:2019jor, Minic:2020oho})
should be extended to obtain precise collider and astrophysical observational signatures. In the absence of any clear trace of new physics at the LHC, it is imperative to focus on very subtle signatures, for example, 
like what these models offer in terms of the geometry and topology of the particle showers and jets.    

Alternatively, spacetime could be constructed directly from information. 
Namely, it was shown in  \cite{Powers:2021rfg}
that one can derive rules of angular momentum addition in quantum mechanics along with Clebsch-Gordan coefficients from first principles just by studying correlations 
in the set of all binary sequences. Non-determinism, superposition, and interference naturally arise by restricting information about the sequences to the counts that specify correlations.  
It is interesting that  one of the counts qualifies as a metric, which gives hope that models for both matter and spacetime may be supported by a single formalism. 
This line of research that basically requires no prior assumptions may be very promising \cite{sam}.

\noindent
{\bf Theory and phenomenology of wormholes}: Wormholes, being exact solutions to Einstein's equations, have attracted much attention not only out of pure academic interest, but also as solutions that could indeed describe some exotic objects in nature. 
As pointed out in \cite{Dai:2019mse,Dai:2019nph,Simonetti:2020vhw,Dai:2020rnc,Dai:2018vrw}, it is remarkable that our current and near future astrophysical observations are already reaching the precision sufficient to distinguish wormholes from black holes. 
The probability that some of the black holes we observe today are actually wormholes is remote. 
However, an eventual discovery of a single wormhole in our astrophysical surveys would be nothing short of spectacular, and thus warrants our earnest efforts. 
On the other side, there has been a very interesting development in the context of quantum wormholes, namely the ER=EPR proposal \cite{Maldacena:2013xja}, which ties up quantum entanglement with these classical solutions of  Einstein's equations. It is no less remarkable that the current terrestrial experiments are sufficient to put strong constraints on this idea and highlight the potential problems that must be resolved in order to develop this idea further \cite{Dai:2020ffw}. Our future efforts should be concentrated on phenomenological and observational aspects associated with wormholes on both microscopic and macroscopic scales. Any significant headway in this direction may revolutionize our understanding of the universe.

\end{document}